\documentclass{article}

\usepackage{arxiv}

\usepackage[utf8]{inputenc} 
\usepackage[T1]{fontenc}    
\usepackage{hyperref}       
\usepackage{url}            
\usepackage{booktabs}       
\usepackage{amsfonts}       
\usepackage{nicefrac}       
\usepackage{microtype}      
\usepackage{lipsum}
\usepackage{amsmath}
\usepackage{graphicx}
\usepackage{float}
\usepackage[]{algorithm2e}
\usepackage{natbib}

\title{Deep Haar scattering networks in pattern recognition: a promising approach}

\author{
  Fernando Fernandes Neto \\
  Department of Business, Accounting and Economics\\
  University of S\~{a}o Paulo\\
  Cidade Universit\'{a}ria, 05508-010, S\~{a}o Paulo, Brazil \\
  \texttt{fernando\_fernandes\_neto@usp.br} \\
   \And
  Alemayehu Solomon Admasu \\
  Department of Physics \& Astronomy\\
  Rutgers University\\
  Piscataway, New Jersey, 08854, USA \\
  \texttt{a.solomon@rutgers.edu} \\
   \And
  Rodrigo de Losso \\
  Department of Business, Accounting and Economics\\
  University of S\~{a}o Paulo\\
  Cidade Universit\'{a}ria, 05508-010, S\~{a}o Paulo, Brazil \\
  \texttt{delosso@usp.br} \\
 \And
  Claudio Garcia\\
  Department of Electrical Engineering\\
  University of S\~{a}o Paulo\\
  Cidade Universit\'{a}ria, 05508-010, S\~{a}o Paulo, Brazil \\
  \texttt{clgarcia@lac.usp.br} \\
   \And
  Pedro Delano Cavalcanti \\
  Department of Physics \& Astronomy\\
  Rio de Janeiro State University\\
  R. S\~{a}o Francisco Xavier - 524, 20559-900, Rio de Janeiro, Brazil \\
  \texttt{pedelano@gmail.com} \\
}

\begin{document}
\maketitle

\begin{abstract}
The aim of this paper is to discuss the use of Haar scattering networks, which is a very simple architecture that naturally supports a large number of stacked layers, yet with very few parameters, in a relatively broad set of pattern recognition problems, including regression and classification tasks. This architecture, basically, consists of stacking convolutional filters, that can be thought as a generalization of Haar wavelets, followed by non-linear operators which aim to extract symmetries and invariances that are later fed in a classification/regression algorithm. We show that good results can be obtained with the proposed method for both kind of tasks. We have outperformed the best available algorithms in 4 out of 18 important data classification problems, and have obtained a more robust performance than ARIMA and ETS time series methods in regression problems for data with strong periodicities.
\end{abstract}


\section{Introduction}

Pattern recognition in time-series is a fundamental data analysis type for understanding dynamics in real-world systems. It is common to gather time-series data from a wide range of problems, such as stock market prediction, speech and music recognition, motion capture data and electronic noise data \citep{most_d}. They can also be obtained by means of successive measurements of higher dimensional problems, such as image contours, sequential counts from network nodes and other mathematical objects, as can be seen in \citep{site}. 

Analysis of time-series data has been the subject of active research for decades and many approaches for modeling them have been developed. Traditional methods, for instance  autoregressive models, Linear Dynamical Systems and Hidden Markov Models (HMM) need an experienced modeler to identify and estimate them, besides the fact that they are subject to failures in modeling accurately complex real-world data \citep{most_d}.

To circumvent these limitations, machine learning based methods became an attractive solution to data analysis of this kind, because they can be applied in linear and non-linear systems and are able to extract features (which can also describe system states) in both Euclidean and non-Euclidean domains, allowing a significant performance gain, as can be seen in \citep{vision_cnn}. 

In this context, in order to increase feature extraction capabilities, machine learning methods have become deeper and deeper, where the most prominent deep learning methods are Convolutional Neural Networks (CNNs). They are employed in a wide range of tasks such as text classification, natural language processing, image processing and time-series data modeling \citep{fernando}. 

CNNs basically consist of multiple convolutional filters, that act as trainable layers, stacked on top of each other and usually followed by a non-linear operator and a pooling layer, followed by a classification algorithm on its tip \citep{vision_cnn}.   

It is important to mention that CNNs also overcome a prevalent problem in most artificial neural networks (ANN), which is the lack of understanding of the underlying statistical and geometric features extracted from the analyzed signal, making the comprehension of why an ANN makes a particular decision a difficult task \citep{black_box, geo_deep}. 

In the quest of trying to understand the success of these algorithms, \citep{invariant}, \citep{Mallat2016} and \citep{geo_deep} have identified that symmetries and invariances play a fundamental role in feature extraction, given that  relevant information contained in a wide range of different signals (such as sounds or images) are typically not affected by translations or rotations and are stable to deformations.

Also, \citep{invariant} suggest that less flexible feature extractors can be obtained by means of simple convolutional filters such as wavelets, followed by simple non-linear operators, yet yielding very good results despite its simplicity. The key factor of this architecture is the  preservation of some important properties of the traditional deep networks, while allowing the reduction of the computational complexity.

Complementing this work and dealing with only Haar Scattering Transforms, which are the simplest Scattering Convolutional Transforms, \citep{deep_haar} show that it is possible to solve traditional classification problems, such as digit recognition, with surprisingly greater mathematical/computational simplicity.

In that way, \citep{fernando} extended this work for 1D signal analyses such as time-series data, showing that general-purpose approximator functions can be obtained based on Haar Scattering Networks, where, for demonstration purposes, only simple Ordinary Least Squares (OLS) regressors were used, with the absolute value function as the non-linear operator.

Having contextualized our research, the main idea of the present paper is to extend  \citep{fernando} by feeding extracted features into classifiers and regressors (such as Support Vector Machines (SVM), OLS regressors and Random Forests) in order to classify/forecast different kinds of signals, using different non-linear operators and an optional pooling layer, which extracts statistical properties of the features, allowing a richer mapping, as can be seen in \citep{bagnall2015}.

We intend to demonstrate that using a very simple architecture, with a relatively large number of stacked layers and a very few parameters, can exhibit very good results - even improving some known results about important problems. These results may open ways to the development of new Automatic Machine Learning (AutoML) algorithms, which is a very recent research field, that aims to find the best performing learning algorithm with minimal human intervention, that is, to automate the design choices of the network (such as topology, optimization procedure, regularization, stability methods) by using hyperparameter optimization \citep{automl}, with computational simplicity.
 
\section{Theory}

\subsection{Wavelet Transforms}

Fourier transforms have many applications in science and engineering, and in the realm of time-invariant signals they provide simple and effective answers to most questions. On the other hand, they become very ineffective with non-stationary problems, due to the fact that sine and cosine functions are just localized in terms of their frequency. They are non-localized in time. In order to solve this problem, a viable substitute for Fourier are Wavelet Transforms.

A discrete wavelet transform is a transform whose basis is composed of a family of orthonormal functions $\psi$, called wavelets, allowing to capture both frequency and location (time and space), unlike classical Fourier Transform \citep{book_mallat}. A Haar wavelet is a particular type of wavelet that is used as the orthonormal basis of the Haar Scattering Network. It is defined by a function $\psi$, as follows.

\begin{equation}
\psi(t) =
\begin{cases}
1, &\text{if } 0 < \text{$t$} \leq 1/2 \\
-1, &\text{if } 1/2 < \text{$t$} \leq 1 \\
0, &\text{otherwise.}
\end{cases}
\end{equation}
Its respective scaling function, $\Psi$, is given by:
\begin{equation}
\Psi(t) =
\begin{cases}
1, &\text{if } 0 < \text{$t$} \leq 1 \\
0, &\text{otherwise.}
\end{cases}
\end{equation}
It is possible to derive, from the definition of a Haar wavelet and wavelet transforms, a pair of equations for calculating the coefficients of the Haar Wavelet Transform, as reviewed in \citep{fernando}:
\begin{equation}
\chi_{\omega}(k,n) = 2^{-1/2}(\chi_{\omega}(2k, n+1) + \chi_{\omega}(2k+1, n+1)) \\
\end{equation}
\begin{equation}
X_{\omega}(k,n) = 2^{-1/2}(\chi_{\omega}(2k, n+1) - \chi_{\omega}(2k+1, n+1)) .
\end{equation}

For more information about the mathematical definitions and properties of Haar wavelets transforms, see \citep{book_mallat} and \citep{invariant}.

\subsection{Haar Scattering Networks}

Scattering networks were introduced as convolution networks, computed with iterated wavelet transforms, to obtain invariants which are stable to deformations \citep{group_invariant, deep_haar}.

A Haar Scattering Network was originally defined in \citep{uns_haar} and \citep{deep_haar} by a sequence of layers, which operates over an input positive $d$-dimensional signal $x \in (\mathbb{R}^{d})^{+}$. The general scheme of Haar Scattering Networks is to iteratively extract Wavelet coefficients of the signal and apply a point-wise absolute value operator on them.

As seen on \citep{uns_haar} a Haar scattering is calculated by iteratively applying the following permutation invariant operator:

\begin{equation}
(\alpha, \beta) \rightarrow (\alpha + \beta, |\alpha - \beta|) .
\end{equation} 
The values $\alpha$ and $\beta$ can be recovered by Equations (6) and (7) enabling to reconstruct the whole previous layer values if $\alpha$ and $\beta$ are real positive:
\begin{equation} \label{eq:max}
max(\alpha, \beta) \rightarrow \frac{1}{2}(\alpha + \beta + |\alpha - \beta|)
\end{equation}
\begin{equation} \label{eq:min}
min(\alpha, \beta) \rightarrow \frac{1}{2}(\alpha + \beta - |\alpha - \beta|) .
\end{equation}
The network layers are defined as two-dimensional arrays $S_{j}x(n,q)$ with dimensions $2^{-j}d \cdot 2^{j}$, where $n$ is a node number and $q$ denotes a feature index.

It follows that $S_{j}$ is a permutation invariant operator that acts over a set of nodes calculated in the previous layer with Equations (8) and (9).

\begin{equation}
S_{j+1}x(n, 2q) \rightarrow S_{j}x(a_{n}, 2q) + S_{j}x(b_{n}, 2q)
\end{equation}

\begin{equation}
S_{j+1}x(n, 2q + 1) \rightarrow |S_{j}x(a_{n}, 2q) - S_{j}x(b_{n}, 2q)|
\end{equation}
where $a_{n}$ and $b_{n}$ work as optimizable maps of pairs, dependent on the features extracted.

The iterative extraction of wavelets coefficients of the signal and the application of absolute point-wise operators can be seen in Figure 1.

\begin{figure}[h]
\centering
\includegraphics[width=0.5\textwidth]{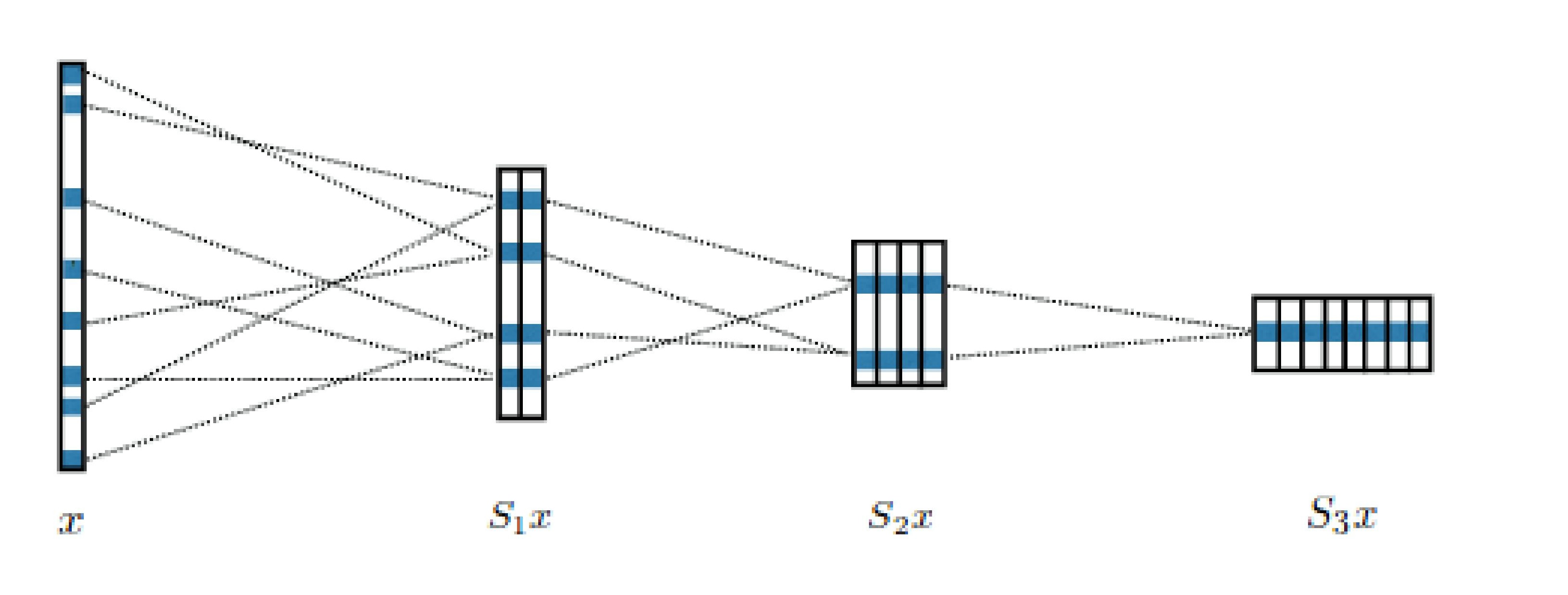}
\caption{A Haar scattering network computes each coefficient of a layer $S_{j+1}x$  by adding or subtracting a pair of coefficients in the previous layer $S_{j}x$ \citep{uns_haar}}.
\end{figure}

Pairing rules ($a_{n}$, $b_{n}$) are optimized, so that we obtain scale and shift parameters, $\sigma$ and $\tau$, respectively. In the current implementation, they act on a signal of length $N$, where $a_{n} = n$ and $b_{n} = (2^{1-j} \cdot N \cdot \sigma) + \tau + n$.

Therefore, these pairing rules differ from the traditional Haar filtering scheme by treating 1D signals as entities that can be represented as graphs, where each node represents a system state, which is directly connected to other states due to their respective multiscale geometric features and invariances, which themselves arise due to possible factors, such as periodicities and trends, that are usually reflected in their spectral or frequency properties.

The main similarities with \citep{fernando} cease here. The key idea of this work is to extend this approach with other non-linear operators (in addition to pointwise absolute value operators) and explore real regression and classification problems using SVMs, OLS regressors as well as random forests.

To achieve this goal, it is also worth mentioning that Haar Scattering Networks, as presented in \citep{uns_haar}, have interesting properties that should be kept when modifying the non-linear operators: the capacity to capture both frequency and location; the convenient information compression initially provided by the absolute value operator and the ability to identify invariants in the data.

\section{Methods}

Following \citep{fernando}, we use the Haar Scattering Network to decompose the original signal in a number of feature-signals, that represent data invariances and symmetries and we feed those features into regressors or classifiers, depending on the type of problem. This architecture is shown in Figure 2.

\begin{figure}[H]
	\centering
	\includegraphics[width=0.5\textwidth]{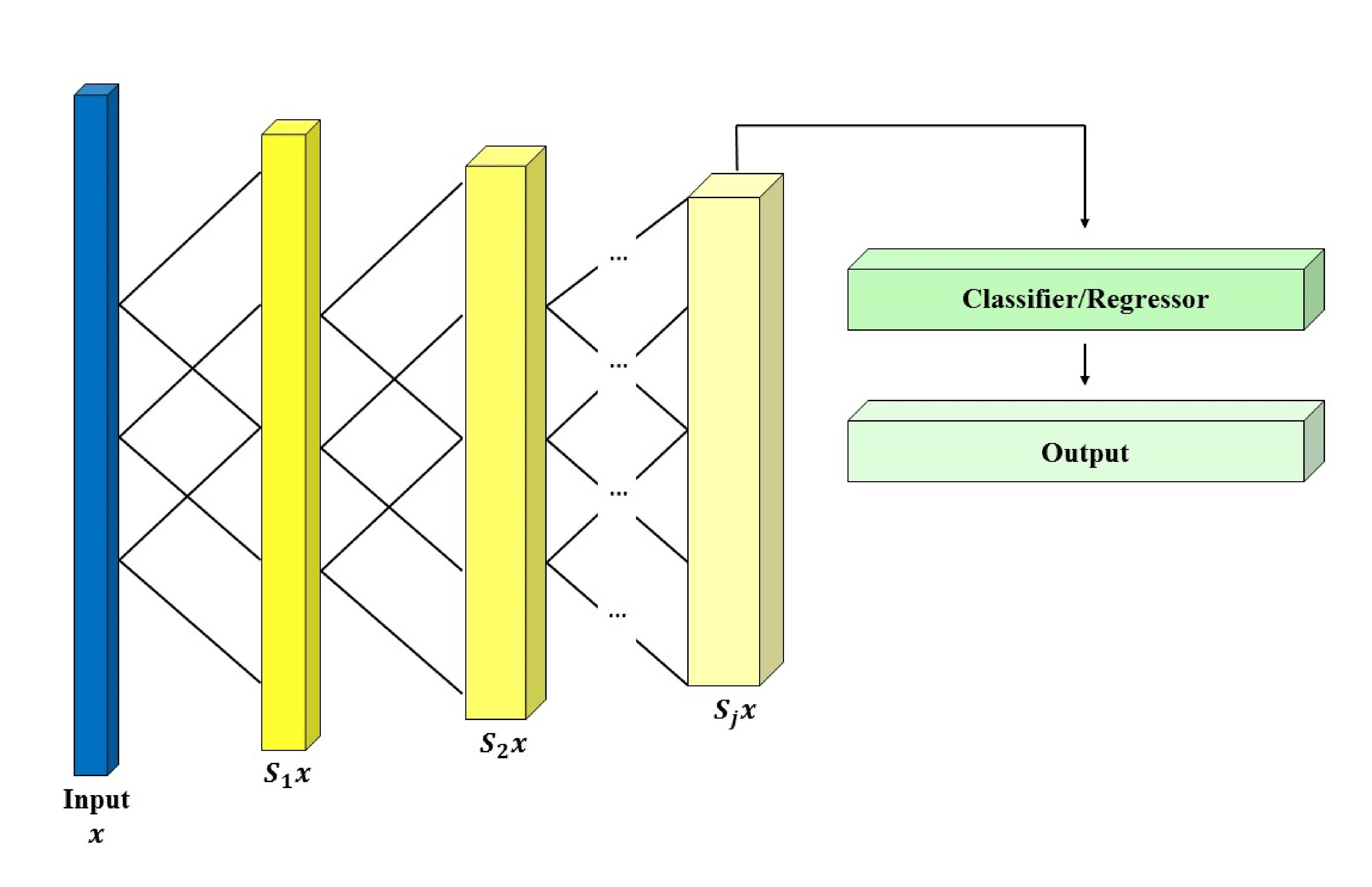}
	\caption{Graphical representation of the original algorithm described on \citep{fernando}.}
\end{figure}

However, in the present paper, we have modified part of the architecture structure aiming to improve its performance. Instead of directly feeding the features into the regressor/classifier, for some classification problems, it is better to introduce a feature transformation layer, which is later fed into the regressor/classifier routines.

The idea of introducing this transformation layer in the original architecture, which was not studied in \citep{fernando} or in \citep{uns_haar}, is to bestow a simple pooling layer - in case of simpler signals - aiming to boost the dimensionality reduction; and to improve the separability of the features in a more robust dimensional space.

When only one property is calculated (such as $max$, $min$ or $mean$), this transformation layer is a pooling layer. On the other hand, when statistical moments, autocorrelation and partial autocorrelation functions are calculated, data can be mapped into other feature spaces, which enhances the quantity of statistical/spectral information available.

While the first approach (simple pooling layer) seems to be an answer to the increasing number of features when the network becomes deeper - counterbalancing the number of features that are fed into the classifier; the second one (spectral features) is an alternative when simple features cannot be linearly separated within simple dimensional spaces by the classifier. 

Also, following the ideas in \citep{residual_cnns}, instead of using only the last layer $S_{j}$ of the Haar Scattering Network, we have observed that results may be significantly better when the inputs of a lower layer are made available to the transformation layer, resembling residual CNNs, of course, depending on the given problem. This is justified by eventual need of maintaining multiscale information about the signal, such as information contained in different frequencies/time scales - in case of time series.

That said, this connectivity increases even more the number of extracted features, pointing towards the necessity of introducing the aforementioned transformation layers: for pooling purposes - in the case of a large number of hidden layers; or for providing a more robust dimensional space, while maintaining relevant multiscale information.

All these changes are summarized in Figure 3 and can be directly compared to the original architecture, which is summarized in Figure 2.

\begin{figure}[H]
	\centering
	\includegraphics[width=0.5\textwidth]{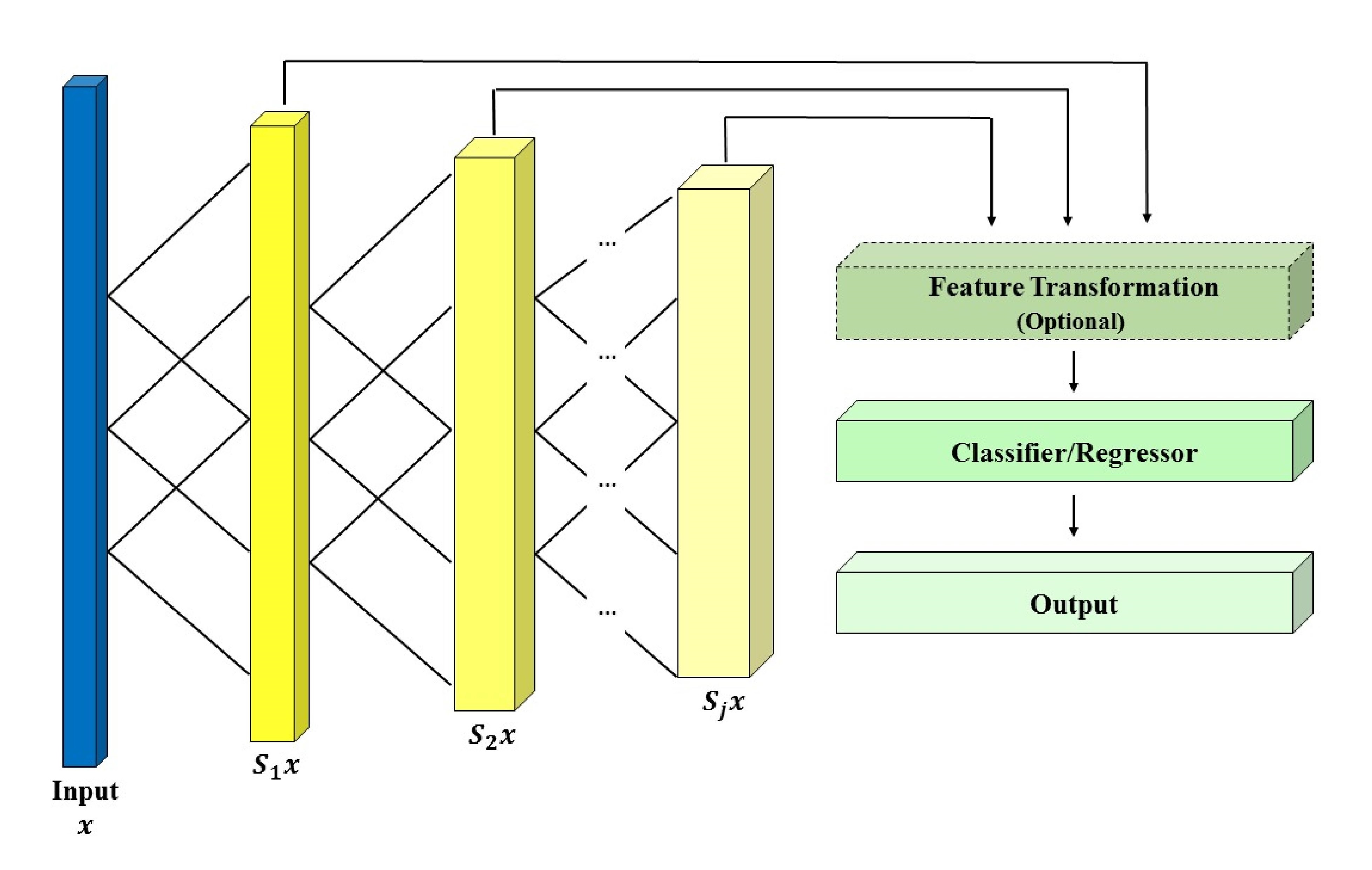}
	\caption{Graphical representation of the algorithm used in this work.}
\end{figure}

Aiming at a better understanding of the algorithm, it is worth writing the processing scheme defined in Figure 1 (which is more compact) as pseudocode, as in Algorithms 1 and 2.

\smallskip

\begin{algorithm}[H]
 \KwData{$S_j$, $\sigma$ and $\tau$;}
 \KwResult{$S_{j+1}$;}
 
 N = length($x$)\;
 $S_{j+1}$ = NewEmptyLayer(N, j)\;
 $N'$ = N * $2^{-j}$\;
 
 \For{$1<= n <= N'$}{
  	 \For{$1<= k <= N'/2$}{
  	$S_{j+1}$[$n$, 2*($k$-1)+1] = $S_{j}$[n, $k$] +  $S_{j}$[$2^{1-j}$*$N$*$\sigma$+$\tau$+$n$, $k$]\;
    }
    
     \For{$1<= k <= N'/2$}{
  	$S_{j+1}$[$n$, 2*$k$] = TF($S_{j}$[n, $k$] - $S_{j}$[$2^{1-j}$*$N$*$\sigma$+$\tau$+$n$, $k$])\;
    }
 }
 \smallskip
 \caption{Internal Layer Processing - Function "HaarLayer"}
\end{algorithm}

\smallskip

Where "TF" stands for "Transfer Function", which is the non-linear operator that should be applied over the differences inherent to the Haar wavelet, possibly different from the absolute value operator proposed in \citep{fernando};

\smallskip

\begin{algorithm}[H]
 \KwData{$x$, $\sigma$ and $\tau$;}
 \KwResult{$S_j x$;}
 $S_0$ = inputLayer($x$)\;
 \For{$0<= j <= L$}{
  	$S_{j+1}$ = HaarLayer($S_j$, $\sigma$, $\tau$)\;
 }
 \smallskip
 \caption{General Haar Network Layer Processing - Function "HaarNetwork"}
\end{algorithm}

\medskip

To obtain the code with residual channels (as shown in Figure 3), the implementation follows the same scheme, but, instead of returning only the last layer, one should create an auxiliary data structure to save the intermediate layers.

Estimations for regression problems are carried out as in Algorithm 3.

\medskip

\begin{algorithm}[H]
 \KwData{$x_t$;}
 \KwResult{Optimal $\sigma$, $\tau$, $R^2$ and $\mathcal{R}$ (Regression Model);}
  $\sigma$ = $\sigma_0$\;
  $\tau$ = $\tau_0$\;
 \While{$|R^2_{k+1} - R^2_{k}| > \varepsilon$}{
  $F(x_{t})$ = HaarNetwork($x_{t}$, $\sigma_k$, $\tau_k$)\;
  $\sigma_{k+1}$, $\tau_{k+1}$, $R^2_{k+1}$, $\mathcal{R}$ = \\
  Optimize\{Regression($F(x_{t})$, $x_t$), $\sigma_k$, $\tau_k$\}\;
 }
 \smallskip
 \caption{General Estimation Procedure for Regression Problems}
\end{algorithm}

\medskip

Estimations for classification tasks are carried out as in Algorithm 4.

\medskip

\begin{algorithm}[H]
 \KwData{$x$, ClassOf($x$);}
 \KwResult{Optimal $\sigma$, $\tau$, $C$ (Success Counts) and \\
 $\mathcal{M}$ (Classification Model);}
 $\sigma$ = $\sigma_0$\;
  $\tau$ = $\tau_0$\;
 \While{$|C_{k+1} - C_{k}| > \varepsilon$}{
  $\mathcal{F}(x)$ = FeatureTransform(HaarNetwork($x$, $\sigma_k$, $\tau_k$))\;
  $\sigma_{k+1}$, $\tau_{k+1}$, $C_{k+1}$, $\mathcal{M}$ = \\
  Optimize\{Classification($\mathcal{F}$($x$), ClassOf($x$)), $\sigma_k$, $\tau_k$\}\;
 }
 \smallskip
 \caption{General Estimation Procedure for Classification Problems}
\end{algorithm}

\medskip

Predicting $k$ steps ahead requires, in terms of regression tasks, that new values of $F(x_{t+k})$ should also be predicted. To accomplish such task, the proposed prediction method for extracted features is to take advantage that they are multiple stochastic processes that preserve different symmetries and invariances (which can be non-linear) from the original signal, forecast $F(x_{t})$ up to $F(x_{t+k})$, by means of Fourier Series (eventually with trends, depending on the signal) and then feed into the estimated regression model $\mathcal{R}$.

On the other hand, classifying out-of-sample observations only requires extracting $\mathcal{F}(x^{*})$, given a new sample $x^{*}$, and calculating the output from the estimated model $\mathcal{M}(\mathcal{F}(x^{*}))$.

Having proposed and described how the algorithm works, an assessment of its performance is needed, in order to compare the proposed method with other well established methods. To accomplish this goal we have proceeded as follows.

For classification problems, we have analyzed 18 datasets from the UEA $\&$ UCR Time Series Classification Repository \citep{site}. They are named as: ``Computers", ``Synthetic Control", ``ECG - 200", ``ECG - 5000", ``Earthquakes", ``Medical Images", ``Phonemes", ``FaceAll", ``Mallat", ``Distant Phalanx Age and Groups", ``Fish", ``Adiac", ``Haptics", ``Insect Wings", ``BeetleFly", ``FordA", ``Chlorine Concentration" and ``Inline Skate". 

For benchmarking purposes, in this kind of task, simple accuracy measures were used (percentage of correct classifications) to assess their performance.

In terms of regression problems, we have analyzed 5 very well known datasets in time series analysis: ``Lung Cancer Deaths - UK", ``Average Monthly Temperature - Nottingham", ``Quarterly Gas Consumption - UK", ``Monthly totals of international airline passengers" and ``Mauna Loa Atmospheric CO2 Concentration". In these cases, out-of-sample $R^2$  measures were used to assess their performance.

The regression and classification algorithms that we have used were Ordinary Least Squares (OLS) estimator, Random Forests \citep{random_forest}, SVMs \citep{svm}, Conditional Trees \citep{conditional_trees} and Recursive Partitioning \citep{recursive_partitioning}. 

The whole implementation of the routines was made possible through the R Statistical Package \citep{R}. Instead of implementing regression and classification algorithms, $rpart$ \citep{rpart}, $ctree$ \citep{ctree}, $libsvm$ \citep{svms} and $randomForest$ \citep{randomforest} were respectively used for Recursive Partitioning, Conditional Trees, SVMs and Random Forests.

\section{Results}

Table 1 shows the results for regression/forecasting tasks, while in Table 2 are shown the results for classification tasks.

In regression/forecasting problems, all results are expressed in terms of the $R^2$ measures, which were calculated using the whole out-of-sample set of observations against the predicted sets (test set), in order to verify the forecasting capabilities of the model as a whole, for each dataset, instead of assessing the capabilities for each observation.

\begin{table*}[!htbp]
\caption{\label{tab1}Summary of the results obtained in the Regression/Forecasting tasks}
\centering
\scalebox{0.9}{
\begin{tabular}{|p{4.5cm}|p{2cm}|p{2cm}|p{2cm}|p{2cm}|p{2.5cm}|}
\hline
Dataset & Out-of-sample Observations & Haar Network $R^2$ & ARIMA Model $R^2$ & 
ETS Model $R^2$ & Automatic Estimated\newline  ARIMA Model\\
\hline
Lung Cancer Deaths - UK &
12 & 0.8920 & 0.7687 & 
0.9509
& ARIMA(2,0,1)\\
\hline
Average Monthly Temperature - Nottingham & 
24 & 
0.9355 & 
0.9243 & 
0.9561
&
ARIMA(5,0,1)\\
\hline
Quarterly Gas Consumption - UK
&
12
&
0.9063
&
--
&
0.7197
& 
ARIMA(2,1,3)\\
\hline
Monthly totals of international airline passengers
& 12
&
0.9360
& 0.9791
&
0.6363 & ARIMA(4,1,3) \\
\hline
Mauna Loa Atmospheric CO$_2$ Concentration & 
24
&
0.8828
&
0.4326
& 0.2709
& ARIMA(3,1,4)\\
\hline
\end{tabular}
}
\end{table*}

\begin{table*}[!htbp]
\caption{\label{tab1}Summary of the results obtained in the Classification tasks}
\centering
\scalebox{0.8}{
\begin{tabular}{|p{2cm}|p{1cm}|p{1.6cm}|p{1.5cm}|p{1.5cm}|p{1.5cm}|p{1cm}|p{1cm}|p{1.4cm}|p{1.7cm}|}
\hline
Dataset & No. of Classes & Haar Network Accuracy & Best Model Accuracy & 
Optim. Procedure & Feature Transf. Layer & T.F. & No. of Layers & Residual Channels & Classif. \newline Algorithm \\
\hline
Computers &
2 &
0.7480 &
0.8 &
Nelder \& Mead &
Maximum Value &
$abs$ &
4 &
Yes &
Random Forest\\
\hline
Synthetic Control & 
6 & 
0.9867 & 
0.9992 & 
Grid Search
&
Median Value &
$tanh$ &
6 &
Yes &
Random Forest\\
\hline
ECG 200 &
2 &
0.91 &
0.8905 &
Nelder \& Mead & 
Median Value &
$tanh$ &
4 &
No &
SVM
\\
\hline
ECG 5000 &
5 &
0.9155 &
0.9461 &
Grid Search & 
Mean Value &
$\sigma$ &
5 &
No &
Conditional Trees\\
\hline
Earthquakes &
2 &
0.7410 &
0.7592 &
Nelder \& Mead & 
Maximum Value &
$abs$ &
2 &
Yes &
Recursive Partitioning\\
\hline
Medical Images &
10 &
0.7118 &
0.7850 &
Grid Search & 
Spectral Properties &
$abs$ &
3 &
Yes &
Random Forest\\
\hline
Phonemes$^*$ &
39 &
0.3387 &
0.3620 &
Nelder \& Mead & 
Spectral Properties &
$tanh$ &
3 &
Yes &
SVM\\
\hline
FaceAll &
14 &
0.9448 &
0.99 &
Nelder \& Mead & 
Spectral Properties &
$\sigma$ &
3 &
Yes &
SVM\\
\hline
Mallat &
8 &
0.8899 &
0.9742 &
Grid Search & 
Maximum Value &
$abs$ &
7 &
Yes &
SVM\\
\hline
Distant Phalanx Age Groups &
3 &
0.7480&
0.8293 &
Grid Search & 
Spectral Properties &
$tanh$ &
2 &
Yes &
Random Forest\\
\hline
Fish &
7 &
0.88 &
0.9742 &
Grid Search & 
Minimum Value &
$\sigma$ &
5 &
No &
SVM\\
\hline
Adiac &
37 &
0.7775 &
0.8098 &
Grid Search & 
Spectral Properties &
$\sigma$ &
2 &
Yes &
SVM\\
\hline
Haptics &
5 &
0.4870 &
0.5096 &
Nelder \& Mead & 
Spectral Properties &
$tanh$ &
3 &
Yes &
SVM\\
\hline
Insect Wings &
11 &
0.6389 &
0.6389 &
Grid Search & 
Maximum Value &
$\sigma$ &
6 &
No &
Random Forest\\
\hline
BeetleFly &
2 &
0.9000 &
0.9485 &
Nelder \& Mead & 
Maximum Value &
$\sigma$ &
3 &
No &
SVM\\
\hline
FordA &
2 &
0.9076 &
0.9654 &
Nelder \& Mead & 
Spectral Properties &
$tanh$ &
3 &
Yes &
SVM\\
\hline
Chlorine Concentration &
3 &
0.8804 &
0.8457 &
Grid Search & 
Mean Value &
$tanh$ &
7 &
Yes &
SVM\\
\hline
Inline Skate$^*$ &
7 &
0.6343 &
0.5525 &
Grid Search & 
Spectral Properties &
$abs$ &
2 &
Yes &
SVM\\
\hline
\end{tabular}
}
\end{table*}

Moreover, in all these problems, the number of layers in the Haar network structure was fixed at 6; absolute value operator was chosen as non-linear operator following \citep{fernando}; all estimation procedures were carried out using the grid search method provided in \citep{NMOF}; the input signals were interpolated using cubic-splines, to provide a larger amount of data to be processed, in order to increase the number of degrees of freedom; and the regression algorithm used was the Ordinary Least Squares (OLS) estimator.

It is also worth mentioning that no Feature Transformation layers were used in these problems, while the residual channels scheme, as in Figure 2, were used in all of them, in order to preserve multiscale information.

For benchmarking purposes, the results obtained using Haar Networks were compared to well known standard methods in time series analysis, as can be seen in \citep{Hamilton}: ARIMA (autoregressive integrated moving average) models and ETS (error, trend, seasonality) models.

Aiming at the estimation of ARIMA models, lags selections were carried out using the $auto.arima$ procedure provided in the R statistical package - which detects the best ARIMA structure using statistical information criteria - for each time series. For the ETS models we used the $ets$ procedure. Both procedures are provided in $forecast$ R package \citep{forecast}.

In Table 2, we show the results for each dataset, specifying: optimization procedure; the type of feature transformation layer; type of non-linear operator (transfer-function); number of Haar Network Layers; if residual channels are present; and, finally, the classification algorithm used.

In addition to that, it is important to notice that 5 different types of feature transformation layers were used: Maximum Value; Minimum Value; Median Value and Mean Value - these four acting as traditional pooling layers; and a different type that calculates some spectral properties of the extracted features. In this case: the first 4 values of the autocorrelation and partial autocorrelation functions - which characterize some of the spectral properties; plus the 4 first statistical moments: mean, variance, kurtosis and skewness.

Three different types of transfer functions (non-linear operators) were also tested: sigmoid function (denoted by $\sigma$); absolute value operator (denoted by $abs$) and hyperbolic tangent function (denoted by $tanh$). Also, in Table 2, dataset names marked with an asterisk indicate whether there is a rebalance in the training set and test set. This procedure was carried out because the original training sets were too small to train an SVM classifier.

\section{Discussion}

From the regression tasks perspective, which can be seen in Table 1, a key finding emerges: the performance of Haar networks in five different well-known datasets, in comparison to well established methods such as ARIMA and ETS models, show that the proposed method is possibly more robust than its counterparts on the average, at least in this class of problems.

Despite not always providing the best performance, the proposed model had an out-of-sample $R^2$ measure above 88\% in all datasets. On the other hand, the performance of ARIMA and ETS models varied in a wide range, from 27\% to 97\%. This highlights how stable the performance delivered by the proposed method is. 

Future research should extend these tests and confirm, to what extent, this method is more robust than its counterparts.

That said, it is important to observe that the results of the experiments in regression tasks found clear support for the fact that the algorithm performs very well in the presence of symmetries and invariances (such as strong seasonal/periodic components) in the data, given that all these time series have linear and non-linear cyclical components.

It is worth noticing that these interesting facts are in line with the perspective of understanding how the proposed algorithm works, in terms of feature extraction, which basically operates by decomposing time series / signals into feature sets that preserve their symmetries.

From the perspective of classification tasks, which can be seen in Table 2, our results cast a new light on how invariances and symmetries play a fundamental role in 1D signal classification.

First, it is important to highlight that, for some of these problems, we have observed that changing the non-linear transfer function from the original absolute value operator as seen in \citep{fernando}, to others, such as $tanh(t)$, made the algorithm perform better. 

Second, we are able to observe empirically, as a clear tradeoff, simple transformation layers are used (to account for dimensionality reduction) for deeper networks (number of layers greater than 3); while, for shallower networks, more complex transformation layers are needed.  

Superior results are seen for "Inline Skate", "Chlorine Concentration" and "ECG 200" datasets, while a negligible improvement is seen in "Insect Wings" dataset. On the remaining datasets, on average, our proposed algorithm is outperformed by 10\% - in terms of the relative performance - by the best algorithms, as compiled by \citep{site}.

Keeping in mind that the proposed algorithm relies on extracting invariances and symmetries and feeding them into an external classifier, this analysis found evidence that this kind of feature plays a fundamental role in 1D signal classification and that a further understanding in conjunction with other well established concepts, such as dynamic time warp, is needed.

It is also worth mentioning that spectral and multiscale features of time-series data can represent some important behavior of the system that are not obvious in the time-series domain. It is possible, for example, using spectral characteristics such as data's frequency and power domain, to extract signal periodicities and reduce data noise.

As already explained before, our approach, an adaptation of \citep{fernando} and \citep{deep_haar}, feeds these spectral and multiscale features to a regression / classification algorithm in order to construct a model of the processes based solely on the sampled data, being an interesting alternative to ARIMA models and to traditional artificial neural networks (ANNs), in a way that additional insights can be retrieved in comparison to these traditional methods.

As can be seen in Figures 4 and 5, our method clearly circumvents the major drawback of  ANNs, which is the fact that they are usually considered "black-boxes", meaning that it is difficult to understand why an ANN makes a particular decision \citep{black_box}. 

In the first case (Fig. 4) it is possible to verify that, naturally, when appropriate transformations are used, clusters arise in different dimensions, allowing desired classification properties. In this case, how myocardial infarction occur, based on the clustering of the features. Dots in red represent normal conditions, turquoise dots are myocardial infarctions and yellow dots represent misclassified cases.

In the second exercise (Fig. 5), it is possible to observe that, if different levels in the original signal are assigned to a specific palette, resembling the colors of a heatmap-like graphic, it is possible to understand how different multiscale properties and their respective signs are linked to the composition of the signal, providing straight-forward information on how these features are correlated to the original process in terms of cycles, trends and irregular components - which is something that ARIMA models and ANN-based models usually do not provide.

\begin{figure}[H]
  \includegraphics[width=70mm]{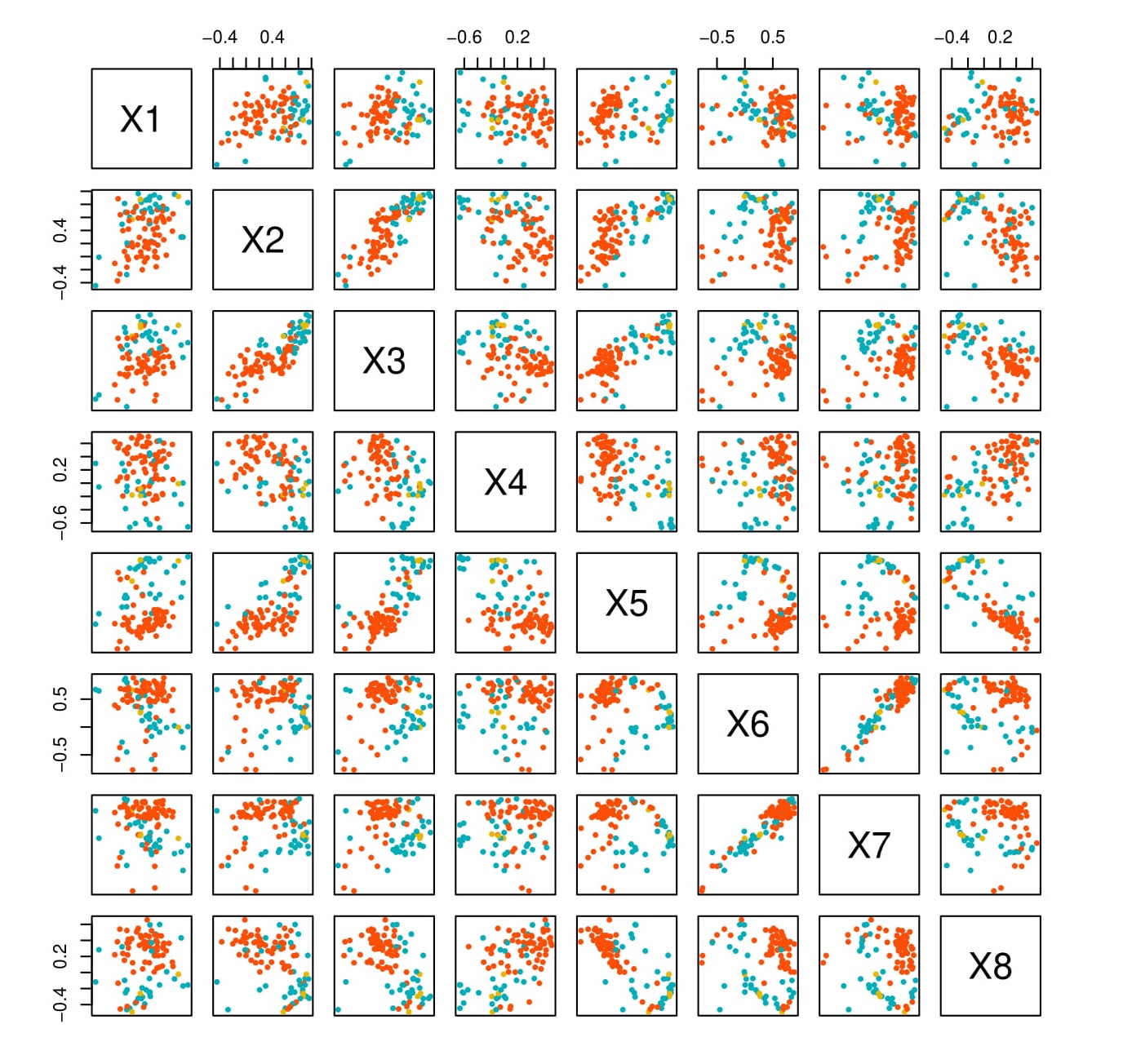}
  \includegraphics[width=70mm]{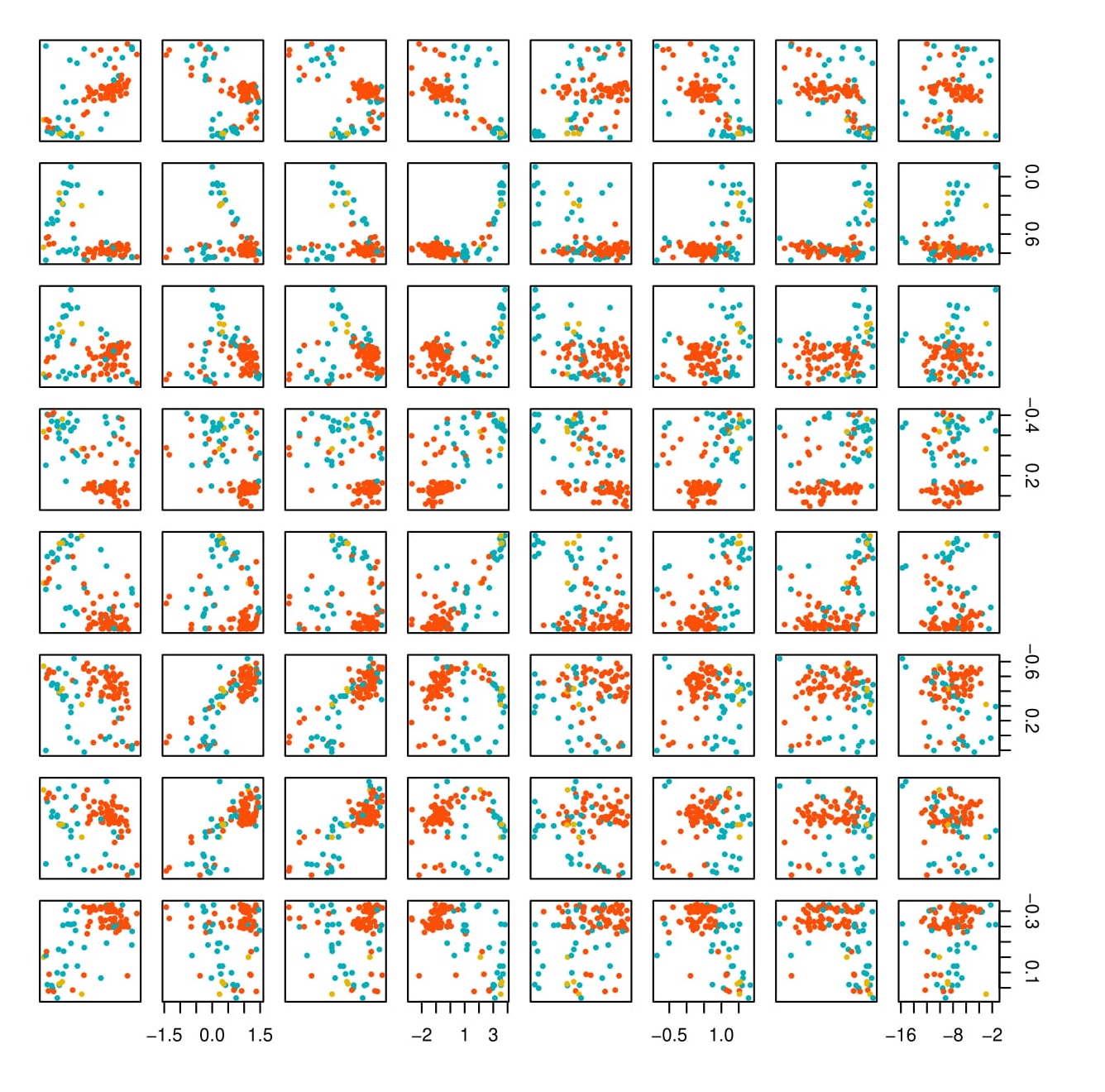}
  \includegraphics[width=70mm]{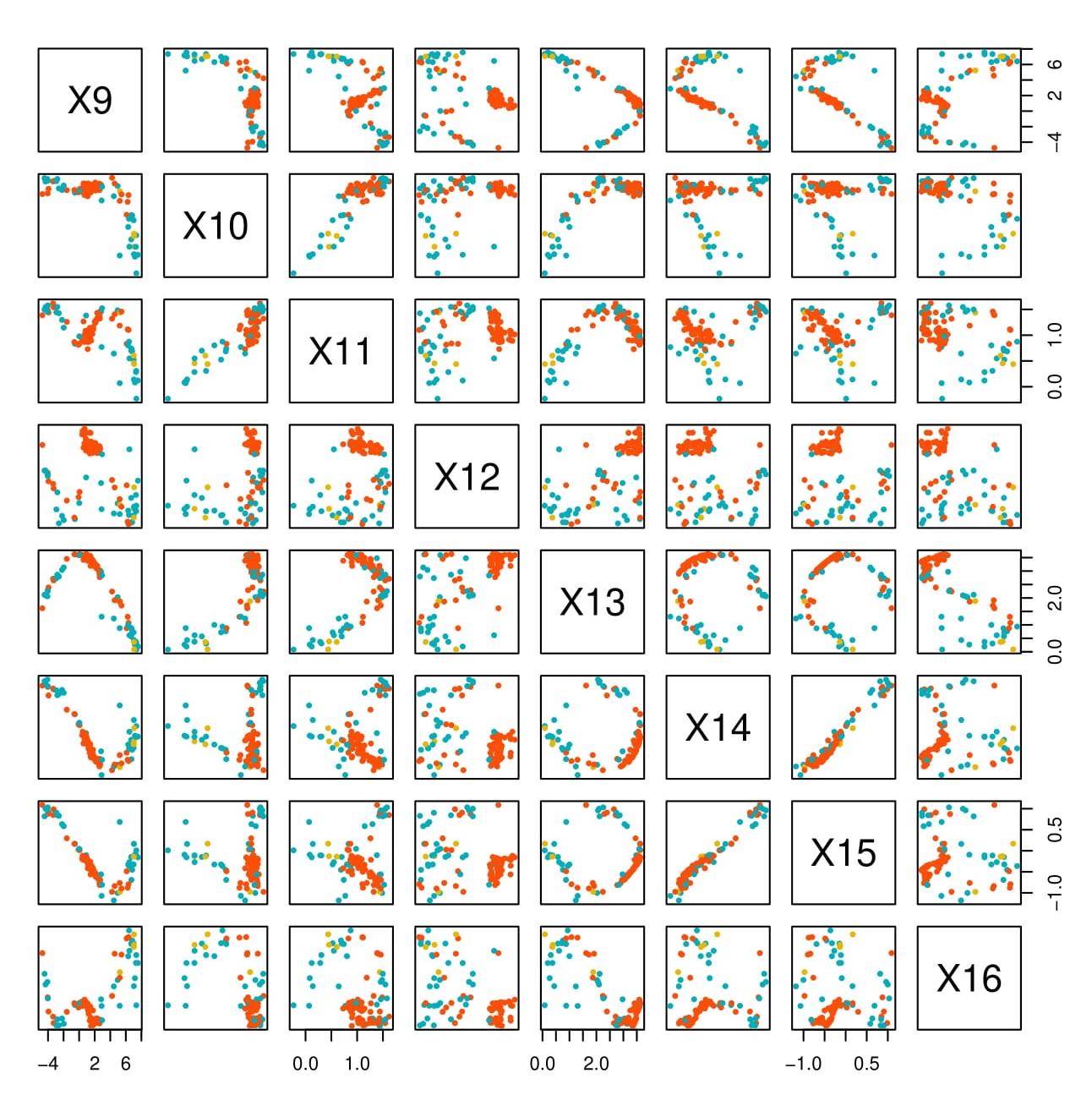}
  \caption{Clustering of features of myocardial infarction in different dimensions}
\end{figure}

\begin{figure}[H]
  \includegraphics[width=75mm]{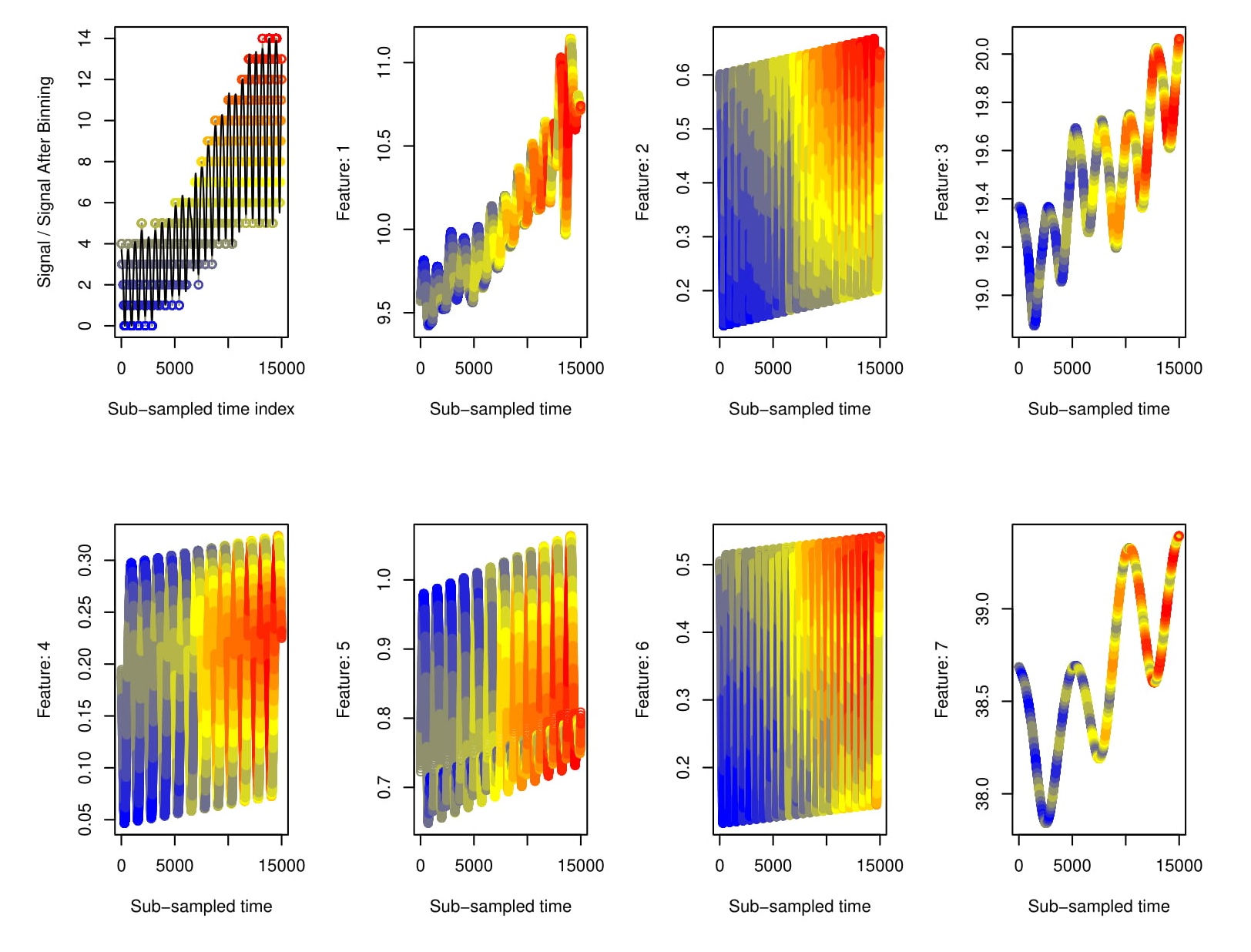}
  \includegraphics[width=75mm]{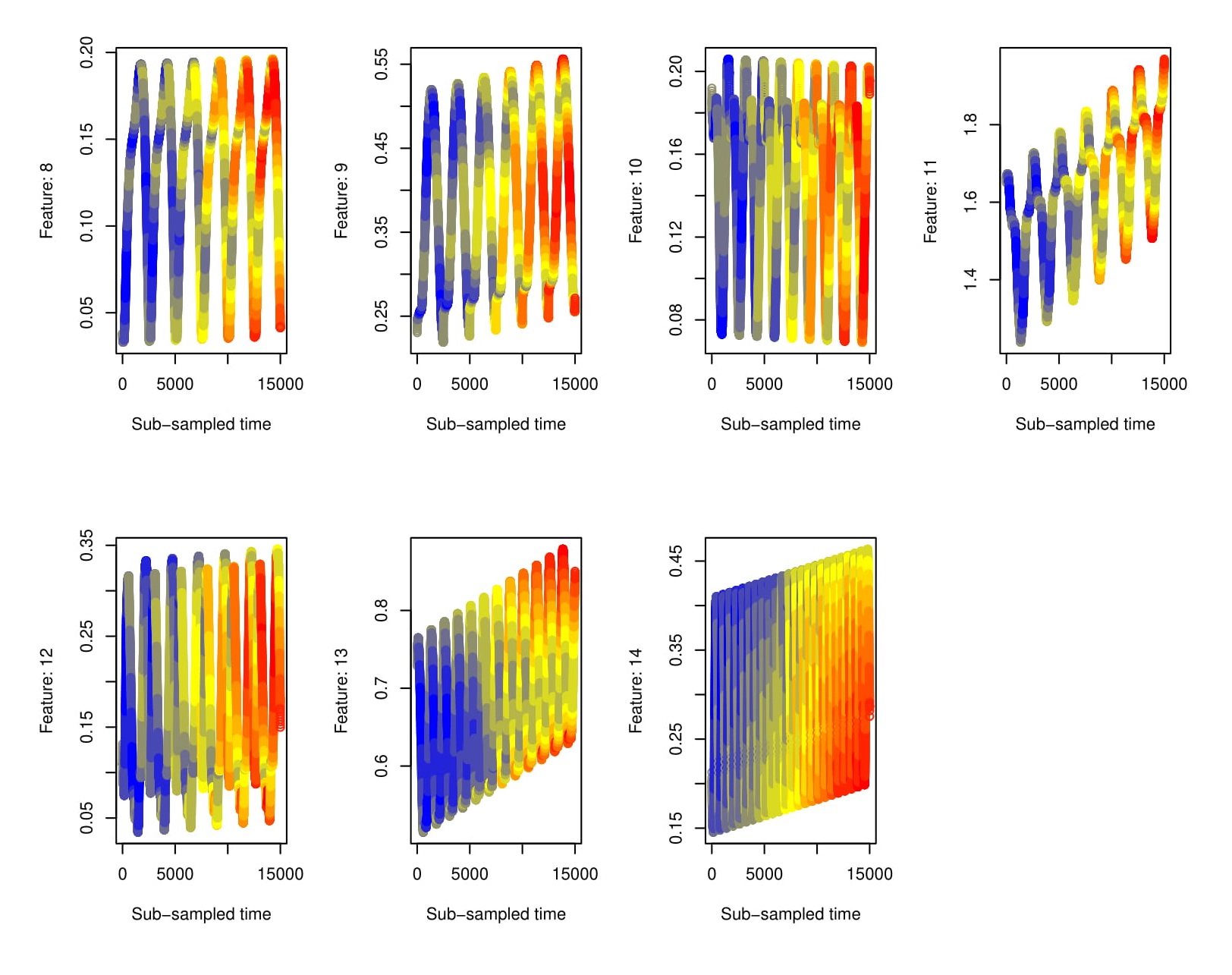}
  \caption{Time Series and its extracted features - UK Gas Consumption Time Series}
\end{figure}

\section{Conclusions}

This paper proposes an extension of the architecture provided in \citep{fernando} - which is basically an adaptation of the architecture developed in \citep{deep_haar} and in \citep{group_invariant}, focusing on 1D signals classification and regression tasks of pattern recognition. This architecture, basically consists of stacking convolutional filters, that can be thought as a generalization of Haar wavelets, followed by non-linear operators, which aim to extract symmetries and invariances that are later fed in a classification/regression algorithm.

We have obtained good results with this simple method, in a wide range of datasets, for both kind of tasks.

Furthermore, despite the fact that dataset descriptions can be found in \citep{just_site}, it is important to highlight and emphasize their important potential real-life applications, such as Myocardial Infarction detection ("ECG - 200"); water quality ("Chlorine Concentration"); Industrial Control Charts ("Synthetic Control") and detection of disease-vectors ("Insect Wings"), such as Aedis Egypt. Thus, the impact of this line of research in pattern recognition is potentially considerable. 

The same conclusions can be drawn for regressions in the presence of strong periodic components, which are interesting to a broader audience of professionals who seek massive automatic modeling, without the need of further investigations, such as economists, engineers and data scientists. As pointed out in the introductory section, the results may also pave the way to the development of new AutoML algorithms, providing functional models with minimal human intervention, given its potential of generalization.

That said, given the considerable flexibility to adapt the architecture to different problems, by means of modifying the non-linear operator, number of layers, feature transformation layer and the classifier itself, it is imperative to extend this research, in further works, to enable a better comprehension of how much can be improved, since the search for the best architecture was primarily hand-made. Other classification algorithms and non-linear operators, which were neglected here, can also be included.

\bibliographystyle{ieeetr}
\bibliography{Main}

\end{document}